\documentclass[12pt]{article} 
\usepackage[utf8]{inputenc} 
\usepackage{geometry}
\usepackage{graphicx} 
\usepackage{amssymb}
\usepackage{amsmath}
\usepackage{cite}
\usepackage{mathrsfs}
\usepackage{booktabs} 
\usepackage{array} 
\usepackage{paralist}
\usepackage{verbatim} 
\usepackage{subfig}
\usepackage{fancyhdr} 
\usepackage{sectsty}
\allsectionsfont{\sffamily\mdseries\upshape} 
\usepackage[nottoc,notlof,notlot]{tocbibind} 
\usepackage[titles,subfigure]{tocloft} 

\title{Construction by similarity method of the fundamental solution of the Dirichlet problem for Keldysh type equation in the half-space.}
\author{\bf Oleg D. Algazin}
\date{Bauman Moscow State Technical University, Moscow, Russia} 

\begin{document}
\maketitle
\thispagestyle{empty}

\begin{abstract}
      For elliptic in the half-space and parabolic degenerating on the boundary equation of  Keldysh type we construct by similarity method the self-similar solution, which is the approximation to the identity in the class of integrable functions. This solution is the fundamental solution of the Dirichlet problem, i.e. the solution of the Dirichlet problem with the Dirac delta-function in the boundary condition. Solution of the Dirichlet problem with an arbitrary function in the boundary condition can be written as the convolution of the function with the fundamental solution of the Dirichlet problem, if  a convolution exists. For a bounded and piecewise continuous boundary function convolution exists and is written in the form of an integral, which gives the classical solution of the Dirichlet problem, and is a generalization of the Poisson integral for the Laplace equation. If the boundary function is a generalized function, the convolution is a generalized solution of the Dirichlet problem.
\end{abstract}


\section*{Introduction}
         In \cite{Kel}  M.V.Keldysh considered the equation
\[
u_{xx}+y^mu_{yy}+a(x,y)u_y+b(x,y)u_x+c(x,y)u=0,~~ (c(x,y)\leq0),\eqno{(1)}
\]
which is elliptic in the upper half-plane $y>0$ and parabolic degenerates on the line $y=0$. For the domain, located in the upper half-plane and having a part of the boundary on  $x$-axis , he proved solvability conditions for the Dirichlet problem with continuous data on the boundary  . These conditions impose restrictions on $m$ and $a(x,0)$. The Dirichlet problem is solvable if:
\[
m<1,~~\text{or}~~  m=1, a(x,0)<1~~\text{or}~~  1<m<2, a(x,0)\leq0  .\eqno{(2)}
\]
In the  plane  Keldysh equation as Tricomi equation belongs to a mixed elliptic-hyperbolic type. Applications such equations are described in \cite{Ber},\cite{Otw}.

       This article discusses the multidimensional generalization of Keldysh equation in a half-space $y>0$
\[
\Delta_xu+y^mu_{yy}+\alpha y^{m-1}u_y-\lambda^2u=0,~~m<2,~~\alpha<1,\eqno{(3)}
\]

where $x=(x_1,\dots,x_n)\in\mathbb{R}^n, u=u(x,y)$~  is a function of variabels~$(x,y)\in\mathbb{R}^{n+1}$, 
\[
\Delta_x=\frac{\partial^2 }{\partial x_1^2}+\dots+\frac{\partial^2 }{\partial x_n^2}
\]
is the Laplace operator on the variable $x$.

Dirichlet condition is given on the boundary of the half-space
\[
                      u(x,0)=\psi(x)   .       \eqno{(4)}
\]
The coefficients of the equation
\[
      a(x,y)=\alpha y^{m-1}   ,b(x,y)=0,c(x,y)=-\lambda^2,    \eqno{(5)}
\]
so Keldysh condition (2) are satisfied.

To build a bounded when $y\to+\infty$ solution of the Dirichlet problem (3), (4) it suffices to find a fundamental solution to the Dirichlet problem, i.e. the bounded for $y\to+\infty$  solution of the problem with the Dirac delta-function in the boundary condition
\[
u(x,0)=\delta(x)   .      \eqno{(6)}
\]
If we denote the fundamental solution of the Dirichlet problem (3), (6) through $P(x,y)$, then by the invariance of the equation (3) with respect to translations along the $x$, the solution of equation (3) with the boundary condition $u(x,0)=\delta(x-\xi)$  is a function $P(x-\xi,y)$ . By virtue of the principle of superposition, which is valid for linear equations, the solution of equation (3) with the boundary condition (4) is the convolution of the fundamental solution of the Dirichlet problem with boundary function
\[
u(x,y)=P(x,y)*\psi(x),
\]

if the convolution exists. In order to find a fundamental solution of the Dirichlet problem, we apply the similarity  method \cite{Sed} - \cite{Blu}, using the symmetry of the equation, that is, the invariance of the equation with respect to certain groups of transformations, what is possible due to the specially chosen coefficients (5).

      This work is a continuation of our work \cite{Alg}, where the case $\alpha = 0, \lambda = 0$ is considered. 

      The fundamental solutions of the Dirichlet problem for the cases   $n=1, \alpha=0,m<2$ and $n=1,\alpha=m<1$ are found by means of Fourier transform  in \cite{Par}.

       In \cite{Gel1}-\cite{Gel3}  fundamental solutions of the Tricomi operator , and in \cite{Che}  fundamental solution of the Keldysh type  operator (that is, solutions of inhomogeneous equations with $\delta$-function in the right-hand side) are found by the similarity method.

\section{Statement of the problem. Approximation to the identity.  }
In the equation (3) and in the boundary condition (6) we make the change of variables
\[
\xi=x,   \eta=\frac{2}{2-m} y^{\frac{2-m}{2}}   . 
\]
We get (again replacing $\xi$ on $x$ and $\eta$ on $y$) the equation
\[
\Delta_xu+u_{yy}+\frac{\beta}{y}u_y-\lambda^2u=0,~\beta=\frac{2\alpha-m}{2-m}<1,~y>0,~x\in\mathbb{R}^n \eqno{(7)}
\]

and the boundary condition
\[
          u(x,0)=\delta(x),~ x\in\mathbb{R}^n.  \eqno{(8)}
\]                          
We seek a solution  $P(x,y)$  of the equation (7), which is a delta-shaped family of functions of   $x\in\mathbb{R}^n$  with the parameter $y\to+0$ or an approximation to the identity.  It is enough to demand the fulfilling  the next properties for any $y>0$
\begin{align*}
1)~&P(x,y)>0,\\
2)~&\int_{\mathbb{R}^n}P(x,y)dx=1,\\
3)~&\forall\delta>0, \lim_{y\to+0} \int_{|x|\ge\delta} P(x,y)=0.
\end{align*}
If these properties are satisfied, then for any bounded piecewise continuous function $f(x)$  there exists a convolution
\[
f(x)*P(x,y)=\int_{\mathbb{R}^n}f(t)P(x-t,y)dt=\int_{\mathbb{R}^n}f(x-t)P(t,y)dt
\]
and in point $x$, in which the function $f(x)$ is continuous
\[
\lim_{y\to+0}f(x)*P(x,y)=f(x).
\]
Indeed, 
\[
\left|f(x)*P(x,y)-f(x) \right|=\left| \int_{\mathbb{R}^n}f(x-t)P(t,y)dt-f(x) \int_{\mathbb{R}^n}P(t,y)dt \right|=
\]
\[
=\left|\int_{\mathbb{R}^n}(f(x-t)-f(x) )P(t,y)dt\right|\le\int_{\mathbb{R}^n}\left|f(x-t)-f(x) \right|P(t,y)dt=
\]
\[
=\int_{|t|<\delta}\left|f(x-t)-f(x) \right|P(t,y)dt+\int_{|t|\ge\delta}\left|f(x-t)-f(x) \right|P(t,y)dt\le
\]
\[
\le\varepsilon\int_{\mathbb{R}^n}P(t,y)dx+2M\int_{|t|\ge\delta}P(t,y)dx<2\varepsilon
\]
for $y$ values sufficiently close to zero, because $\int_{|t|\ge\delta}P(t,y)dx<\frac{\varepsilon}{2M}$, by condition 3) and $|f(x-t)-f(x) |<\varepsilon$ for sufficiently small $\delta$ by the continuity.

     In particular, for an infinitely differentiable finite functions $\varphi(x)$ from the space  $\mathscr{D}(\mathbb{R}^n)$ \cite{Vla} , we have
\[
\lim_{y\to+0}\int_{\mathbb{R}^n}\varphi(x)P(x,y)\,dx= \varphi(0)=\langle\delta(x),\varphi(x) \rangle,
\]
that is, $P(x,y)$ converges to the $\delta$-function  at $y\to+0$ in a space of generalized functions $\mathscr{D'}(\mathbb{R}^n)$  . By virtue of the continuity of convolution, for the generalized function $f(x)\in\mathscr{D'}(\mathbb{R}^n)$, for which a convolution with $P(x,y)$ exists, the equality holds
\[
\lim_{y\to+0}f(x)*P(x,y)=f(x)*\delta
(x)=f(x).
\]

Condition 3) can be replaced by condition 
\[
3^*)~ \forall\delta>0, \lim_{y\to+0} \sup_{|x|\ge\delta}P(x,y)=0.
\] 
Condition 3) follows from condition  $3^*)$. Indeed,
\[
\int_{|x|\ge\delta}P(x,y)\,dx=\int_{\delta\le|x|\le\Delta}P(x,y)\,dx+\int_{|x|\ge\Delta}P(x,y)\,dx\le
\]
\[
\le \int_{\delta\le|x|\le\Delta}\sup_{|x|\ge\delta}P(x,y)\,dx+\int_{|x|\ge\Delta}P(x,y)\,dx.
\]
The first integral is small with a small $y$  (condition  $3^*$), and the second integral is small for sufficiently large $\Delta$ as the remainder of a convergent integral (condition 2).

      Condition 2) can be replaced by condition
\[
2^*) ~ \lim_{y\to+0} \int_{\mathbb{R}^n}P(x,y)\,dx=1  .
\]
Indeed, denote
\[
\int_{\mathbb{R}^n}P(x,y)\,dx=g(y)>0  , ~  \lim_{y\to+0}g(y)=1,~P^* (x,y)=\frac{P(x,y)}{g(y)}   .
\]
Then $P^* (x,y)$ satisfies the conditions 1) -3) and therefore is an approximation to the identity, and
\[
\lim_{y\to+0}P^*(x,y)=\lim_{y\to+0} P(x,y).
\]

\section{The case $\lambda = 0$}
Consider the Dirichlet problem
\[
\Delta_x u+u_{yy}+\frac{\beta}{y} u_y=0,~\beta<1,~    y>0,~x\in\mathbb{R}^n  ,       \eqno{(9)}
\]
\[         
                                          u(x,0)=\delta(x),~   x\in\mathbb{R}^n. \eqno{(10)}                          
\]              
Equation (9) is invariant under dilations
\[
\bar x=tx,~\bar y=ty,~\bar u=t^k u,~t>0,~k \text{ is any number.}
\]
Therefore, we consider the solutions of equation (9), which are invariant under this group of transformations, that is the solutions which are homogeneous functions $u(tx,ty)=t^{-k}u(x,y)$.  Passing to the limit as $y\to+0$, we get  $u(tx,0)=t^{-k} u(x,0)$, but because of the boundary conditions (10)  $u(x,0)=\delta(x)$ and, therefore, we have the equality $\delta(tx)=t^{-k}\delta(x)$. It follows that $k=n$, since $\delta(x)$ in $\mathbb{R}^n$   is a homogeneous function of degree $–n$ \cite{Vla},\cite{GeS}
\[
\delta(tx)=t^{-n}\delta(x).
\]
We will seek a self-similar solution of equation (9) as a homogeneous function of degree  $-n$
\[
u=\frac{1}{y^n}\varphi\left(\frac{|x|}{y}\right),~|x|=r.
\]
Equation (9) can be written as
\[
u_{rr}+\frac{n-1}{r}u_r+u_{yy}+\frac{\beta}{y}u_y=0.
\]
Substituting in it the function
\[
u=\frac{1}{y^n}\varphi\left(\frac{r}{y}\right)~  \text{and denoting}~  \frac{r}{y}=\xi,
\]
we obtain the ordinary differential equation for the function $\varphi(\xi)$
\[
(1+\xi^2 ) \varphi'' (\xi)+\left(\frac{n-1}{\xi}+(2n+2-\beta)\xi\right) \varphi' (\xi)
+(n^2+n-\beta n)\varphi(\xi)=0.
\]
We make the change of variables
\[
1+\xi^2=\eta ,
\]
and  denoting the function $\varphi(\sqrt{\eta-1})$  through $\bar \varphi(\eta)$, we obtain the equation
\[
\eta (1-\eta )\bar \varphi'' (\eta )+\left(\frac{n}{2}+\frac{3-\beta}{2}-\left(n+\frac{3-\beta}{2}\right)\eta\right)\bar  \varphi' (\eta )
-\left(\frac{n^2}{4}+\frac{n(1-\beta)}{4}\right)\bar \varphi(\eta )=0.
\]
It is a hypergeometric equation
\[
\eta(1-\eta)\bar \varphi''(\eta)+\left(c-\eta(a+b+1)\right)\bar \varphi'(\eta)-ab\bar \varphi(\eta)=0,
\]
where
\[
a= \frac n2,~b=\frac n2+\frac{1-\beta}2,~c=1+\frac n2+\frac{1-\beta}2.
\]
Its general solution has the form
\[
\bar \varphi(\eta)=C_1F(a,b;c;\eta)+C_2\eta^{1-c}F(b-c+1,a-c+1;2-c;\eta),
\]
where $F$ is hypergeometric function. Because  $b-c+1=0$, then
\[
F(b-c+1,a-c+1;2-c;\eta)=1.
\]
We take a particular solution
\[
\bar \varphi(\eta)=C_n\eta^{1-c}=\frac{C_n}{\eta^{n/2+(1-\beta)/2}}.
\]
Returning to the old variables, we obtain
\[
\varphi(\xi)=\frac{C_n}{(1+\xi^2)^{n/2+(1-\beta)/2}},~~u=\frac 1{y^n}\varphi\left(\frac ry\right)=\frac{C_ny^{1-\beta}}{(y^2+r^2)^{n/2+(1-\beta)/2}}.
\]
We choose a such constant $C_n$  , that the integral of the function $\varphi(|x|),|x|=r$  on the whole space $\mathbb{ R}^n$  will be equal to the unity. We get, denoting $\sigma_{n-1}$ area of the unit sphere in $\mathbb{ R}^n$ ,
\[
C_n^{-1}=\sigma_{n-1}\int_0^{\infty}\varphi(r)r^{n-1}\,dr=\sigma_{n-1}\int_0^{\infty}\frac{r^{n-1}\,dr}{(1+r^2)^{n/2+(1-\beta)/2}}=
\]
\[
=\frac{\sigma_{n-1}}2\int_0^{\infty}\frac{t^{n/2-1}}{(1+t)^{n/2+(1-\beta)/2}}\,dt=\frac{\sigma_{n-1}}2\frac{\Gamma\left(\frac n2\right)\Gamma\left(\frac{1-\beta}2\right)}{\Gamma\left(\frac n2+\frac{1-\beta}2\right)}=\frac{\pi^{n/2}\Gamma\left(\frac{1-\beta}2\right)}{\Gamma\left(\frac n2+\frac{1-\beta}2\right)}.
\]
The solution of the equation (9) is a function
\[
P_0(x,y)=C_n\frac{y^{1-\beta}}{(y^2+|x|^2)^{n/2+(1-\beta)/2}},~\text{where}~C_n=\frac{\Gamma\left(\frac n2+\frac{1-\beta}2\right)}{\pi^{n/2}\Gamma\left(\frac{1-\beta}2\right)}.\eqno{(11)}
\]
This function satisfies the conditions 1)-3) and hence is a fundamental solution of the Dirichlet problem for the equation (9). Solution of the Dirichlet problem for the equation (9) with an arbitrary function at the boundary condition $u(x,0)=\psi(x)$  will be the convolution (if the convolution exists)
\[
                      u(x,y)=\psi(x)*P_0 (x,y). \eqno{(12)}
\]   
If  $\psi(x)$ is a generalized function and convolution (12) exists, it gives a generalized solution of the Dirichlet problem:
\[
\lim_{y\to+0}u(x,y)=\psi(x)~\text{in}~   \mathscr{D}' (\mathbb{R}^n ),
\]  
\[
\text{that is}~ \forall\varphi(x)\in\mathscr{D}(\mathbb{R}^n )~ \lim_{y\to+0}\int_{\mathbb{R}^n}u(x,y)\varphi(x)\,dx=\langle\psi(x),\varphi(x)\rangle.
\]  
       If  $\psi(x)$ is a piecewise continuous bounded function, then the convolution (12) exists and is recorded in the form of integral
\[
u(x,y)=C_n\int_{\mathbb{R}^n}\frac{y^{1-\beta}\psi(t)\,dt}{(y^2+|x-t|^2)^{n/2+(1-\beta)/2}},~ C_n=\frac{\Gamma\left(\frac n2+\frac{1-\beta}2\right)}{\pi^{n/2}\Gamma\left(\frac{1-\beta}2\right)},
\]                                          
which gives the classical solution of the Dirichlet problem, that is, at each point of continuity of the function $\psi(x),~ \lim_{y\to+0}u(x,y)=\psi(x)$.

Replacing in (11) $y$ on $2y^{(2-m)/2}/(2-m)$  and $\beta$  on  $(2\alpha-m)/(2-m)$, we obtain the fundamental solution of the Dirichlet problem
\[
Q_0(x,y)=C_n^*\frac{y^{1-\alpha}}{(y^{2-m}+(2-m)^2|x|/4)^{n/2+(1-\alpha)/(2-m)}},
\]
\[
C_n^*=\frac{(2-m)^n\Gamma(n/2+(1-\alpha)/(2-m))}{2^n\pi^{n/2}\Gamma((1-\alpha)/(2-m))},
\]
for equation (3), $\lambda=0$ :
\[
\Delta_xu+y^mu_{yy}+\alpha y^{m-1}u_y=0,~y>0,~x\in\mathbb{R}^n,~m<2,~\alpha<1.
\]

\section{The case $\lambda\not=0$}
We will seek a fundamental solution of the Dirichlet problem to the equation
\[
\Delta_xu=u_{yy}+\frac{\beta}yu_y-\lambda^2u=0\eqno{(13)}
\]
in the same form, that a found fundamental solution of the Dirichlet problem (11) for the equation (9) :
\[
u=y^{1-\beta}f(y^2+|x|^2),~~|x|=r.
\]
Substituting the function $u=y^{1-\beta}f(y^2+r^2)$ into the equation
\[
u_{rr}+\frac{n-1}ru_r+\frac{\beta}yu_y-\lambda^2u=0,~~\beta<1,
\]
and denoting $\xi=y^2+r^2$, we obtain   the ordinary differential equation for the function $f(\xi)$
\[
4\xi f''(\xi)+(6+2n-2\beta)f'(\xi)-\lambda^2f(\xi)=0.
\]
This equation reduces to the Bessel equation and has a solution, tending to zero at infinity \cite{Kam}
\[
f(\xi)=C_n\frac{K_{\nu}\left(\lambda\sqrt{\xi}\right)}{\left(\sqrt{\xi}\right)^{\nu}},~~\nu=\frac{1+n-\beta}2,
\]
where  $K_{\nu}$  is a  MacDonald function. Returning to the old variables, we get
\[
u(x,y)=C_n\frac{y^{1-\beta}K_{\nu}\left(\lambda\sqrt{r^2+y^2}\right)}{\left(\sqrt{r^2+y^2}\right)^{\nu}},~r=|x|,~\nu=\frac{1+n-\beta}2.
\]
We choose a such constant $C_n$  , that the integral of the function $u(x,y)$  on the whole space $\mathbb{R}^n$ tends to the unity when $y\to+0$. Turning to spherical coordinates, we get \cite{Rys}
\[
\int_{\mathbb{R}^n}u(x,y)\,dx=C_n\sigma_{n-1}y^{1-\beta}\int_0^{\infty}\frac{r^{n-1}K_{\nu}\left(\lambda\sqrt{r^2+y^2}\right)}{\left(\sqrt{r^2+y^2}\right)^{\nu}}\,dr=
\]
\[
=C_n\left(\frac{2\pi}{\lambda}\right)^{n/2}y^{(1-\beta)/2}K_{(1-\beta)/2}\left(\lambda y\right).
\]
Considering the asymptotic behavior of the Macdonald function
\[
K_{\mu}(z)\sim\frac12\Gamma(\mu)\left(\frac z2\right)^{-\mu},~z\to0,
\]
we have
\[
\lim_{y\to+0}y^{(1-\beta)/2}K_{(1-\beta)/2}(\lambda y)=\frac{\Gamma((1-\beta)/2)}{2^{(1+\beta)/2}\lambda^{(1-\beta)/2}}
\]
and
\[
C_n=\frac{\lambda^{\nu}}{2^{\nu-1}\pi^{n/2}\Gamma((1-\beta)/2)}.
\]
Fundamental solution of the Dirichlet problem for the equation (12) is a function
\[
P_{\lambda}(x,y)=\frac{\lambda^{\nu}}{2^{\nu-1}\pi^{n/2}\Gamma((1-\beta)/2)}\frac{y^{1-\beta}K_{\nu}\left(\lambda\sqrt{|x|^2+y^2}\right)}{\left(\sqrt{|x|^2+y^2}\right)^{\nu}},
\]
because it satisfies the conditions 1),2*), 3), which determines the approximation to the identity.
     Passing here to the limit when $\lambda$ tends to zero and taking into account the asymptotics of the MacDonald function, we obtain the fundamental solution (11) of the Dirichlet problem for the equation (9)
\[
P_0(x,y)=\frac{\Gamma(\nu)}{\pi^{n/2}\Gamma\left((1-\beta)/2\right)}\frac{y^{1-\beta}}{\left(|x|^2+y^2\right)^{\nu}},~\nu=\frac{1+n-\beta}2.
\]
Replacing in $P_{\lambda}(x,y)$ $y$ on $2y^{(2-m)/2}/(2-m)$  and $\beta$  on  $(2\alpha-m)/(2-m)$, we obtain the fundamental solution of the Dirichlet problem for the equation (3).

\textbf{Example.} 
Consider the Dirichlet problem
\[
u_{xx}+u_{yy}=0,~~y>0,~~x\in\mathbb{R}^n,
\]
\[
u(x,0)=x_+^{-3/2},
\]
\[
u(x,y)~\text{is bounded when}~y\to+\infty.
\]
Here
\[
x_+^{-3/2}=
\begin{cases}
x^{-3/2},~&x>0\\
0,&x<0
\end{cases}
\]
This function has a nonintegrable singularity in zero, so it generates a singular generalized function, operating on the basic functions $\varphi(x)\in\mathscr{D}$ according to the rule \cite{GeS}
\[
\langle x_+^{-3/2},\varphi(x)\rangle=\int_0^{\infty}\frac{\varphi(x)-\varphi(0)}{x^{3/2}}\,dx.
\]
The Laplace equation is obtained from equation (3) when 
$ n=1,m=0,\alpha=0,\lambda=0$ and the fundamental solution of the Dirichlet problem for him is the Poisson kernel
\[
P(x,y)=\frac y{\pi(x^2+y^2)}.
\]
Solution to the Dirichlet problem is a convolution
\[
u(x,y)=x_+^{-3/2}*P(x,y)
\]
To calculate this convolution, we first consider the Dirichlet problem with boundary condition $u(x,0)=-2x_+^{-1/2}$. This function is locally integrable and generates a regular generalized function, generalized derivative of which is equal to $x_+^{-3/2}$\cite{GeS}. For the function $-2x_+^{-1/2}$  convolution with the Poisson kernel is recorded by the integral
\[
-2x_+^{-1/2}*P(x,y)=-\frac{2y}{\pi}\int_0^{\infty}\frac{dt}{\sqrt{t}((x-t)^2+y^2)}=
\]
\[
=-\frac{\sqrt{2}y}{\sqrt{x^2+y^2}\sqrt{\sqrt{x^2+y^2}-x}}.
\]
Now, using the differentiation property of convolution, we find
\[
u(x,y)=x_+^{-3/2}*P(x,y)=\frac{\partial}{\partial x}\left(-2x_+^{-1/2}*P(x,y)\right)=
\]
\[
=\frac{y\left(3x\sqrt{x^2+y^2}-3x^2-y^2\right)}{\sqrt{2}\left(x^2+y^2\right)^{3/2}\left(\sqrt{x^2+y^2}-x\right)^{3/2}}.
\]
For an arbitrary generalized function in the boundary condition of the Dirichlet problem, for which exists a convolution with the fundamental solution of the Dirichlet problem, we can  guarantee the convergence of convolution to the boundary function when $y\to+0$ only  in the weak sense, that is convergence in $\mathscr{D}'$. But for this function exists the usual limit 
\[
\lim_{y\to+0}u(x,y)=0~\text{if}~ x<0~\text{and}~  \lim_{y\to+0}u(x,y)=\frac 1{x^{3/2}}~\text{if}~x>0.
\]

\section*{                                             Conclusion}
    
For a multidimensional elliptic equation in half-space with parabolic degeneracy at the boundary, which is a generalization of the Keldysh equation we found the fundamental solution of the Dirichlet problem by the  similarity method. The solution of the Dirichlet problem with an arbitrary boundary function is written as a convolution of this function with the fundamental solution of the Dirichlet problem.


\begin{thebibliography}{99}
\bibitem{Kel}
Keldysh M.V. On some cases of degenerate elliptic equations on the boundary of a domain ,Doklady Acad. Nauk USSR.  Vol. 77 (1951), 181-183.
\bibitem{Ber}
Bers L. Mathematical Aspects of Subsonic and Transonic Gas Dynamics, Surveys Appl. Math. 3, Wiley, New York, 1958.
\bibitem{Otw}
Otway T.H. Dirichlet Problem for Elliptic-Hyperbolic Equations of Keldych Type. Springer-Verlag, Berlin, Heidelberg, 2012. 
\bibitem{Sed}
Sedov L. I. Similarity and dimensional methods in mechanics. – CRC press, 1993.
\bibitem{Bar}
Barenblatt G.I. Scaling, Self-Similarity, and Intermediate Asymptotics, Cambridge University Press, 2002.
\bibitem{Ovs}
Ovsyannikov L. V. Group Analysis of Differential Equations, Academic Press, New York.1982.
\bibitem{Ibr}
Ibragimov N.H. Group analysis of ordinary differential equations and the invariance principle in mathematical physics // Russian mathematical surveys. 1992, t.47, issue 4. P.83-144.(in Russian)
\bibitem{Blu}
Bluman G.W., Cole J.D. Similarity Methods for Differential Equations. Springer-Verlag, New-York,  Heidelberg, Berlin, 1974. 333 P.
\bibitem{Alg}
Algazin O.D. Exact solution to the Dirichlet Problem for Degenerating on the Boundary Elliptic equation of Tricomi — Keldysh type in the half-space. Vestn. Mosk. Gos. Tekh. Univ. im.N.E. Baumana, Estestv. Nauki [Herald of the Bauman Moscow State Tech. Univ., Nat. Sci.],2016, no. 5, pp. 4–17. 
    DOI: 10.18698/1812-3368-2016-5-4-17 , E-print:  arXiv:1603.05760
\bibitem{Par}
Parasyuk L.S. , Parasyuk I.L. The properties of fundamental solutions of the basic boundary value problems for some second order differential equations of mixed type, Ukrainska akademia drukarstva.  Naukovi zapysky. 1999, No.1. pp. 126-129 (in Ukrainian)
\bibitem{Gel1}
Barros-Neto J., Gelfand I.M. Fundamental solutions for the Tricomi Operator , Duke Math. J. 98 (3) (1999), 465–483.
\bibitem{Gel2}
Barros-Neto J., Gelfand I.M. Fundamental solutions for the Tricomi operator, II,Duke Math. J. 2002. 111 (3),  561–584.
\bibitem{Gel3}
Barros-Neto J., Gelfand I.M. Fundamental solutions for the Tricomi operator, III,Duke Math. J. 2005. 128 (1),  119–140.
\bibitem{Che}
Chen ShuXing. The fundamental solution of the Keldysh type operator // Science in China Series A: Mathematics Sep. 2009. Vol. 52. No. 9. P. 1829-1843. DOI:10.1007/s11425-009-0069-8
\bibitem{Vla}
Vladimirov V.S. Generalized Functions in Mathematical Physics, Moscow, Nauka Publ., 1979 (in Russian).
\bibitem{GeS}
Gelfand I.M. and  Shilov G.E., Generalized Functions, Publish info New York, Academic Press, 1964.
\bibitem{Kam}
Kamke E. Handbook of ordinary differential equations . Chelsea Publ.  1976.
\bibitem{Rys}
Ryshik I.M., Gradstein I.S. Tables of Integrals, Series, and Products, Academic Press, New York, 2007.

\end{thebibliography}
\end{document}